\documentclass[twocolumn, superscriptaddress, a4paper]{revtex4}
\usepackage{amsfonts}
\usepackage{amsmath}
\usepackage{amssymb}
\usepackage{dcolumn}
\usepackage{bm}
\usepackage{graphicx}
\usepackage{geometry}
\providecommand{\U}[1]{\protect\rule{.1in}{.1in}}
\usepackage[english]{babel}

\begin{document}
\title{Quantum dynamics on a lossy non-Hermitian lattice}

\author{Li Wang}
\email{liwangiphy@sxu.edu.cn}
\affiliation{Institute of Theoretical Physics, State Key Laboratory of Quantum Optics and Quantum Optics Devices, Collaborative Innovation Center of Extreme Optics, Shanxi University, Taiyuan 030006, P. R. China}

\author{Qing Liu}
\affiliation{Institute of Theoretical Physics, State Key Laboratory of Quantum Optics and Quantum Optics Devices, Collaborative Innovation Center of Extreme Optics, Shanxi University, Taiyuan 030006, P. R. China}

\author{Yunbo Zhang}
\affiliation{Key Laboratory of Optical Field Manipulation of Zhejiang Province and Physics Department of Zhejiang Sci-Tech University, Hangzhou 310018, P. R. China}

\begin{abstract}
We investigate quantum dynamics of a quantum walker on a finite bipartite non-Hermitian lattice, in which the particle can leak out with certain rate whenever it visits one of the two sublattices. Quantum walker initially located on one of the non-leaky sites  will finally totally disappear after a length of evolution time and the distribution of decay probability on each unit cell is obtained. In one regime, the resultant distribution shows an expected decreasing behavior as the distance from the initial site increases. However, in the other regime, we find that the resultant distribution of local decay probability is very counterintuitive, in which a relatively high population of decay probability appears on the edge unit cell which is the farthest from the starting point of the quantum walker. We then analyze the energy spectrum of the non-Hermitian lattice with pure loss, and find that the intriguing behavior of the resultant decay probability distribution is intimately related to the existence and specific property of edge states, which are topologically protected and can be well predicted by the non-Bloch winding number. The exotic dynamics may be observed experimentally with arrays of coupled resonator optical waveguides.
\end{abstract}

\maketitle

\bigskip

\section{Introduction}  \label{intro}
Quantum walk\cite{Aharonov,kempe}, originated as a quantum generalization
of classical random walk, has now become a versatile quantum-simulation scheme which has been experimentally implemented in many physical settings\cite{jbbook}, such as optical resonators\cite{Bouwmeester}, cold atoms\cite{Karski,greiner15}, superconducting qubits\cite{rudneryaoprl,pansci,panprl}, single photons\cite{Broome,xueprevival}, trapped ions\cite{Schmitz}, coupled waveguide arrays\cite{Sansoni} and nuclear magnetic resonance\cite{Duj}.
For standard Hermitian systems, quantum walk has been proposed to detect topological phases\cite{kitagawaT,kitagawaE,rudneryaoprx}. And those fundamental effects of quantum statistics\cite{Bordone,qinxz}, interactions\cite{qinxz, wangepjd, Lahini, wangpra, liwang}, disorders\cite{Yiny,BordoneA}, defects\cite{zjpra,zjsci},  and hopping modulations\cite{zjsci,liwang,liwang15,Kraus,liwang17} on the dynamics of quantum walkers have also been intensively investigated.

Recently, non-Hermitian physics\cite{bender98,bender02,bender07,nhqm,uedareview,fu,floscilate,chenwinding,chenclass,
zhai,budich,hujp,yoshida,dujf,lvrong,songz,ueda18,
duanlm,uedaprx,PTlattice,kouspr,harari,bandres,kante,obusePT,wangdamping, chenpan,zoller,cirac} has been attracting more and more research attentions, since gain and loss is usually natural and unavoidable in many real systems, such as coupled quantum dots\cite{rudner09}, optical waveguides\cite{rudner15}, optical lattices\cite{kitagawabloch,takahashi1,takahashi2,takahashi3} and exciton-polariton condensates\cite{amo,gtg}.
In this context, the central concept of bulk-boundary correspondence which was developed for Hermitian systems is carefully examined and reconsidered in many concrete non-Hermitian models\cite{uedareview,martinez,leykam,xiong,torres,songzb,rosenow,regnault,yiweiprb}. Anomalous zero-energy edge state is found in a non-Hermitian lattice which is described by a defective Hamiltonian\cite{lee}. The concept of generalized Brillouin zone (GBZ) is proposed and a non-Bloch band theory for non-Hermitian systems is established for one-dimensional tight-binding models\cite{shunyu,murakami,kunst,wangreal,fang,hujpa,a2219}. With the aid of non-Bloch winding number, the bulk-boundary correspondence for non-Hermitian systems is restored. Concurrently, the study on quantum walk has also been extended to non-Hermitian systems. Quantum dynamics of non-Hermitian system is believed to be quite different from that of standard Hermitian case. And topological transitions in the bulk have already been observed for open systems by implementing non-unitary quantum walk experimentally\cite{rudner15,xuepeng17119,xuepengbbc,xuepengEdge}.

In this work, we consider a non-Hermitian quantum walk on a finite bipartite lattice in which there exists equal loss on each site of one sublattice. Whenever the quantum walker resides on one of the lossy sites, it will leak out at a rate that is determined by the imaginary part of the on-site potential. As time elapses, the quantum walker initially localized on one of the non-decaying sites will completely disappear from the bipartite lattice eventually. Given the ability to record the position from where decay occurs, one may routinely obtain the resultant decay probability distribution. Intuitively, one may expect the decay probability on each unit cell decreases as its distance from the starting point of the quantum walker increases since each unit cell has a leaky site with equal decay strength. Surprisingly, our numerical simulation displays a very counterintuitive distribution of the decay probability in one parametric region, while the intuitive picture described above shows in the rest region. A conspicuous population of decay probability appears on the edge unit cell which is the farthest from the initial position of the quantum walker, while there exists a lattice region with quite low population between the edge unit cell and the starting point.
We analyze the energy spectrum of the finite bipartite non-Hermitian lattice with open boundary condition. It is shown that the exotic distribution of decay probability is closely related to the existence and specific property of the edge states, which can be well predicted by the non-Bloch winding number\cite{shunyu,murakami}.

The paper is organized as follows. In Sect. \ref{secii}, we introduce the bipartite non-Hermitian model with pure loss. And detailed description of the quantum walk scheme is also addressed. In Sect. \ref{pmdistri}, concrete numerical simulations are implemented for a finite non-Hermitian lattice with open boundary condition. Corresponding distributions of the local decay probability obtained numerically are shown for several typical choices of the model parameters. We then compute the band structure of the finite bipartite lattice with open boundary condition in Sect. \ref{En}. Portraits of the intriguing edge states are pictured therein. And with a constant potential shift, our model is transformed into a model possessing balanced gain and loss. Accordingly, both the Bloch and non-Bloch topological invariants which are vital to bulk-boundary correspondence are calculated. Finally, a summary along with brief discussion is given in Sect. \ref{summary}.

\begin{figure}[btp]
\includegraphics[width=0.45\textwidth]{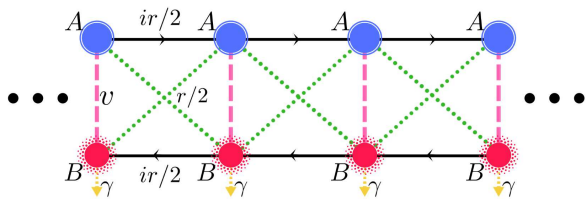}
\caption{(Color online)
Schematic figure of the tight-binding non-Hermitian lattice. Each unit cell contains two sites, A and B. Decay with rate $\gamma$ occurs on each site of the sublattice B. The arrow denotes the phase direction.}
\label{fig1}
\end{figure}

\section{Model and method\label{secii}}

We investigate continuous-time quantum walks on a finite one-dimensional bipartite lattice of length $L$ with pure loss, which is pictured in Fig. \ref{fig1}.  This tight-binding model can be well described by a non-Hermitian Hamiltonian $H$, which reads
\begin{eqnarray}
&H=\sum_{m} \left [ -\frac{i\gamma}{2} \hat{c}^{\dagger}_{m,B} \hat{c}_{m,B}+ v \left( \hat{c}^{\dagger}_{m,A} \hat{c}_{m,B} + \hat{c}^{\dagger}_{m,B} \hat{c}_{m,A} \right) \right.  \nonumber  \\
&+ \frac{ir}{2} \left(  \hat{c}^{\dagger}_{m+1,A} \hat{c}_{m,A} -\hat{c}^{\dagger}_{m,A} \hat{c}_{m+1,A}   \right)    \nonumber \\
&-\frac{ir}{2}   \left(  \hat{c}^{\dagger}_{m+1,B} \hat{c}_{m,B}-\hat{c}^{\dagger}_{m,B} \hat{c}_{m+1,B} \right)  \nonumber \\
&\;\;\;+\frac{r}{2} \left( \hat{c}^{\dagger}_{m+1,A} \hat{c}_{m,B}+\hat{c}^{\dagger}_{m,B}  \hat{c}_{m+1,A} \right)       \;\;\;\;\;
 \nonumber  \\
&\;\;\;+\frac{r}{2} \left. \left( \hat{c}^{\dagger}_{m+1,B} \hat{c}_{m,A}+\hat{c}^{\dagger}_{m,A}  \hat{c}_{m+1,B} \right) \;\; \right],
\label{H}
\end{eqnarray}
with $m$ being the unit cell index. $\hat{c}_{m,A}^{\dagger}$ ($\hat{c}^{\dagger}_{m,B}$) is the creation operator of particles on sublattice $A$ ($B$). $\gamma$ dictates the non-Hermitian loss on each site of $B$ sublattice. And $v$, $r$ denote the intracell and intercell Hermitian hopping, respectively. $\gamma$, $r$, and $v$ are taken to be real, and henceforth.

Accordingly, the dynamics of quantum walker in state $\left|\psi\right>$ dwelling on such a bipartite lattice with long-range hopping obeys the following equations of motion
\begin{eqnarray}
i \dot{\psi}_m^A&=v\psi_m^B\!+\!\frac{ir}{2}\!\left( \psi_{m\!-\!1}^A \! - \! \psi_{m\!+\!1}^A   \right)+\frac{r}{2}\!\left(\psi_{m\!-\!1}^B\!+\!\psi_{m\!+\!1}^B \right),  \nonumber \\
i \dot{\psi}_m^B&=-\frac{i\gamma}{2}\psi_m^{B}+v\psi_m^A-\frac{ir}{2}\left( \psi_{m-1}^B - \psi_{m+1}^B \right) \quad\;\;\quad \nonumber \\
&+\frac{r}{2}\left(\psi_{m-1}^A+\psi_{m+1}^A \right), \quad\;\, \label{tdseq}
\end{eqnarray}
in which the planck constant $\hbar$ is set to be 1. $\psi_m^A=\left< mA| \psi \right>$ and $\psi_m^B=\left< mB | \psi\right>$ are the amplitudes of the quantum walker on site $m$ of sublattice $A$ and $B$, respectively. Since the Hamiltonian (\ref{H}) with pure loss is non-Hermitian in genuine, the norm of the state $\left|\psi\right>$ of the quantum walker will decay in a manner as following,
\begin{eqnarray}
\frac{d}{dt}\left<\psi|\psi\right>=i\left<\psi \left|\left(\hat{H}^{\dagger}-\hat{H}\right)\right|\psi\right>=-\sum_m\gamma\left|\psi_m^B\right|^2.
\label{normdecay}
\end{eqnarray}

Suppose the quantum walker is initially prepared on site $o$ of the sublattice $A$ at time $t=0$, then the initial state $\left|\psi(0)\right>$ of the quantum walker is given by following amplitudes
\begin{eqnarray}
\psi_{m}^{A}(0)=\delta_{mo}, \qquad \psi_{m}^{B}(0)=0.
\label{initialstate}
\end{eqnarray}
For time $t>0$, the quantum walker will move freely on the bipartite lattice according to the equations of motion (\ref{tdseq}).
Due to the existence of pure loss in Hamiltonian (\ref{H}), whenever the quantum walker visits the sites of sublattice $B$, it will leak out with a rate $\gamma$ according to equation (\ref{normdecay}). As $t\rightarrow \infty $, the probability of the quantum walker dwelling on the lattice decreases to be zero. Given the ability to detect the position of the site from where the probability of the quantum walker leaks out, one can
obtain the local decay probability $P_m$ on each leaky unit cell $m$. According to equation (\ref{normdecay}), we have
\begin{eqnarray}
P_m=\int_{0}^{\infty} \gamma \left|\psi_m^B(t)\right|^2 dt,
\label{pm}
\end{eqnarray}
with $\sum_{m}P_m=1$. For the initial state given by equation (\ref{initialstate}), by numerically solving the equations of motion (\ref{tdseq}) with open boundary condition, we can acquire the amplitude $\psi_m^B(t)$ of the quantum walker at any time $t>0$. Throughout this article, we are interested in the resultant distribution of local decay probability $P_m$ among the whole lattice obtained after the quantum walker completely decayed.

\section{Distribution of the local decay probability $P_m$} \label{pmdistri}

We investigate dissipative quantum walks on a finite lattice with $L$ unit cells and under open boundary condition. Without loss of generality, the size of the lattice is taken to be $L=51$. The quantum walker is set out from the non-leaky site of unit cell $o$ in the bulk. As mentioned in Sect. \ref{secii}, the bipartite lattice sketched in Fig. \ref{fig1} is a system with pure loss on each $B$ site, one may immediately has an intuitive picture in mind that the local decay probability $P_m$ shrinks quickly as the distance of the unit cell $m$ from the starting point of the quantum walker increases since the decay strength on each $B$ site is equal. The underlying reason for this is obvious. First come, first served. Quantum walker visits the nearby unit cells first, then more probability leaks out there. Because, as time elapses, the remaining part of the norm of the quantum walker state $\left|\psi(t)\right>$ becomes smaller and smaller. However, direct numerical simulations present intriguing distributions of the local decay probability $P_m$. The picture turns out to be quite counterintuitive where a relatively high population of the local decay probability on the edge unit cell occurs in the resultant distribution. This is very surprising since the edge unit cell is the farthest from the initial position of the quantum walker.

\begin{figure}[t]
\includegraphics[width=0.46\textwidth]{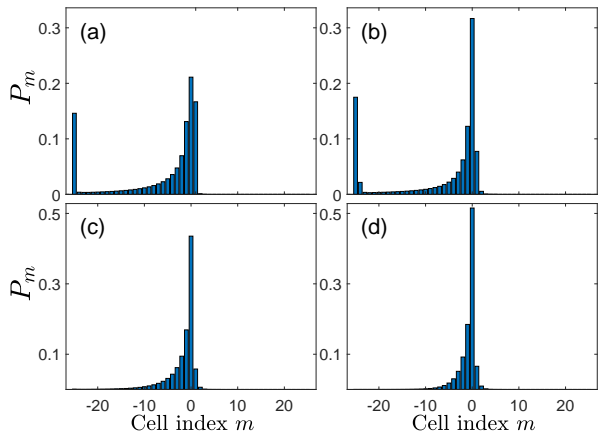}
\caption{(Color online)
Resultant distributions of the local decay probability $P_m$ obtained at the end of the non-Hermitian quantum walks on a finite bipartite lattice.  The intracell hopping $v$ takes positive values, with (a) $v=0.3$, (b) $v=0.5$, (c) $v=0.7$, (d) $v=0.9$. The lattice consists of $L=51$ unit cells with $r=0.5$ and the decay strength $\gamma=1$.}
\label{fig2}
\end{figure}

In Fig. \ref{fig2}(a-d), we simulate the non-Hermitian quantum walk for positive intracell hopping $v$ by numerically solving the equations of motion (\ref{tdseq}). The resultant distributions of local decay probability $P_m$ among the whole lattice are shown for the intracell hopping $v$ taking values $0.3$, $0.5$, $0.7$, $0.9$. And the decay strength is set to be $\gamma=1$, the intercell hopping strength to be $r=0.5$. As shown in Fig. \ref{fig2}(a-d), distributions of the local decay probability are all asymmetric. The quantum walker initiated from the center unit cell $o$ tends to move to the left of the bipartite lattice for positive intracell hopping. And more surprising is that for $v=0.3$ and $v=0.5$ as shown in Fig. \ref{fig2}(a-b), an impressive portion of the probability decays from the left edge unit cell which is the farthest one from the unit cell $o$. Besides, the intuitive picture previously mentioned also shows up,  which is shown in Fig. \ref{fig2}(c-d) for the intracell hopping $v=0.7$ and $v=0.9$. As the distance of the unit cell $m$ from the center unit cell $o$ increases, portion of the probability that leaks out from $m$ becomes smaller and smaller.

\begin{figure}[t]
\includegraphics[width=0.46\textwidth]{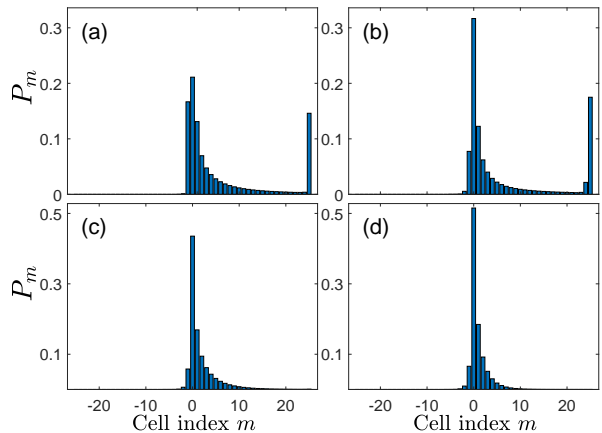}
\caption{(Color online)
Resultant distributions of the local decay probability $P_m$ obtained at the end of non-Hermitian quantum walks on a finite bipartite lattice with $L=51$ unit cells for negative intracell hoppings $v$.  (a) $v=-0.3$, (b) $v=-0.5$, (c) $v=-0.7$, (d) $v=-0.9$. The decay strength $\gamma=1$ and $r=0.5$.}
\label{fig3}
\end{figure}

We then simulate the non-Hermitian quantum walk for negative intracell hopping $v$ with other parameters the same as the positive case above. Details of the distributions of local decay probability $P_m$ are shown in Fig. \ref{fig3}(a-d). Similar to the case of positive $v$, the resultant distributions are also asymmetric. However, in this case the quantum walker has a tendency to go to the opposite direction. Namely, most of the probability of the quantum walker flows to the right side of the bipartite lattice and leaks out there subsequently. Also, as shown in Fig. \ref{fig3}(a-b), a conspicuous population of the decay probability appears on the rightmost unit cell for intracell hopping $v=-0.3$ and $v=-0.5$. And as the strength of the intracell hopping increases, for the cases $v=-0.7$ and $v=-0.9$ as shown in Fig. \ref{fig3}(c-d), the expected distribution of local decay probability $P_m$ is restored again.

\begin{figure}[bp]
\includegraphics[width=0.35\textwidth]{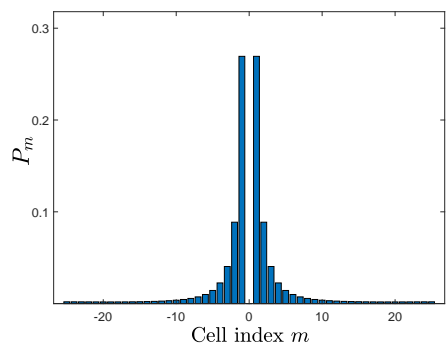}
\caption{(Color online)
Symmetric distribution of local decay probability $P_m$ at the end of the non-Hermitian quantum walk on a finite bipartite lattice with $L=51$ unit cells for intracell hopping $v=0$, decay strength $\gamma=1$ and $r=0.5$.
}
\label{fig4}
\end{figure}

When it comes to the bipartite lattice with zero intracell hopping, there is no direct particle exchange between the two sites within the same unit cell. The quantum walker set out from the central unit cell $o$ will 
preferentially go to lattice sites of nearby two unit cells $o-1$ and $o+1$ rather than the lossy site $B$ of unit cell $o$. Therefore, little probability leaks out from the starting point of the quantum walker. Indeed, this is the case revealed by numerical simulation of a quantum walk in the lossy non-Hermitian lattice with intracell hopping $v=0$, see Fig. \ref{fig4}. In contrast to the counterintuitive cases with finite strength of intracell hopping as shown in Figs. \ref{fig2} and \ref{fig3}, the distribution of local decay probability $P_m$ is nearly symmetric among the whole lattice.

Interestingly, the quantum walk dynamics demonstrated by the numerical simulations above seems quite like a quantum switch. And apparently, by modulating the strength of the intracell hopping $v$, the quantum walker could be regulated at will to reach the left edge unit cell, the right edge unit cell, or none of them with an impressive portion of the probability. This mechanism may have potential applications in the designing of micro-architectures for quantum information and quantum computing in future.

\section{Energy spectrum of the lossy bipartite lattice} \label{En}

To gain a deep insight into the exotic dynamics shown above, in this section we turn to analyze the band structure of the finite bipartite non-Hermitian lattice with open boundary condition in real space. Varying the strength of intracell hopping $v$, the corresponding Hamiltonian matrices of equation (\ref{H}) are numerically diagonalized and the energy spectrum is obtained.

\begin{figure}[t]
\includegraphics[width=0.47\textwidth]{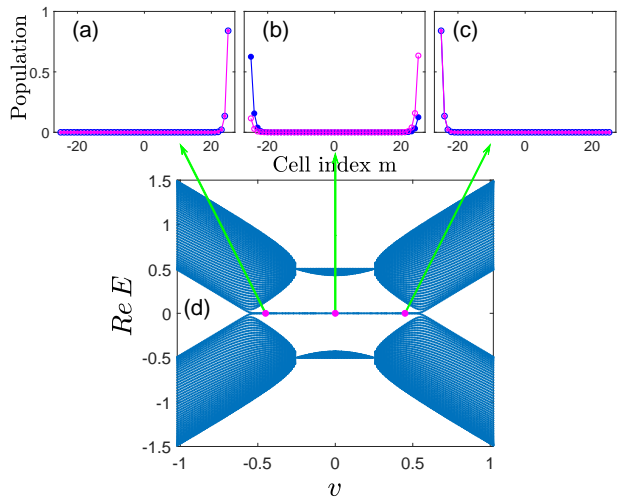}
\caption{(Color online)
Energy spectrum versus intracell hopping $v$ of the finite bipartite non-Hermitian lattice with pure loss under open boundary condition. The lattice size is $L=51$ (unit cell) with the decay rate $\gamma=1$ and intercell hopping $r=0.5$. (a-c) Three typical profiles of edge states. (d) Real part of the single-particle energy spectrum versus intracell hopping $v$.
}
\label{fig5}
\end{figure}

As the Hamiltonian in equation (\ref{H}) is non-Hermitian in nature, the eigenenergies are complex in general. The real parts of the eigenenergies versus the strength of intracell hopping $v$ are plotted in Fig. \ref{fig5}d with $L=51$ unit cells, dissipation strength $\gamma=1$ and $r=0.5$. Apparently, zero modes show up in the real part of the single-particle energy spectrum.
We plot three typical profiles of edge states in Fig. \ref{fig5}(a-c). Interestingly, for positive intracell hopping, the two edge states are both localized on the leftmost unit cell and the curves of their probability population among the whole lattice coincide, see Fig. \ref{fig5}c. Similarly, for negative intracell hopping, the probability population curves also coincide. However, the probability of them are both localized on the rightmost cell in this case,as shown in Fig. \ref{fig5}a. And specially, for zero intracell hopping, the curves of the probability population of the two edge states no longer coincide. From Fig. \ref{fig5}b, we can see that one of the two edge states is localized on the left edge while the other is localized on the opposite side.

\begin{figure}[b]
\includegraphics[width=0.47\textwidth]{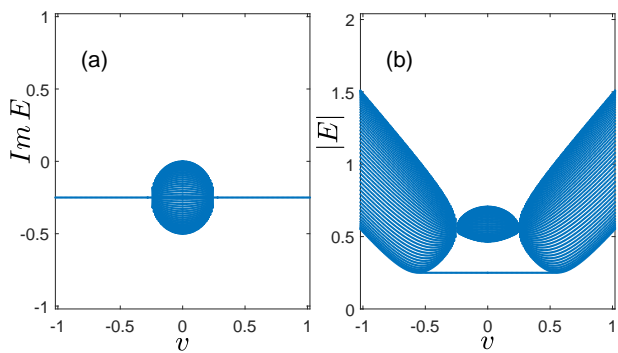}
\caption{(Color online)
Energy spectrum versus intracell hopping $v$ of the finite bipartite non-Hermitian lattice with pure loss under open boundary condition. The lattice size is $L=51$ (unit cell) with the decay rate $\gamma=1$ and intercell hopping $r=0.5$. (a) Imaginary part of single-particle energy spectrum versus intracell hopping $v$.
(b) $|E|$ as a function of the intracell hopping $v$.}
\label{fig6}
\end{figure}

Correspondingly, the imaginary part of the open-boundary energy spectrum is shown in Fig. \ref{fig6}a. It is shown that imaginary parts of the eigenenergies are all located in the lower half plane. This manifests that the eigenstates are going to decay with time. And we plot $|E|$ as function of the intracell hopping $v$ in Fig. \ref{fig6}b where a length of straight line which is well separated from the spectrum bulk of $|E|$ is also shown. These eigenenergies correspond to the edge states.

To investigate topological properties of the model equation (\ref{H}), it is beneficial to pass to the momentum space by fourier transformation. Straightforwardly, the Bloch Hamiltonian is
\begin{eqnarray}
H_k=h_x \sigma_x + \left( h_z+\frac{i\gamma}{4} \right ) \sigma_z - \frac{i\gamma}{4}I
\label{Hbloch}
\end{eqnarray}
where $h_x=v+r \cos k$, $h_z=r \sin k$, $I$ is identity matrix and $\sigma_{x,y,z}$ are pauli matrices.
By compensating an overall gain term, we arrive at $H'_k=h_x \sigma_x + (h_z+i\gamma/4 ) \sigma_z$.

\begin{figure}[t]
\includegraphics[width=0.36\textwidth]{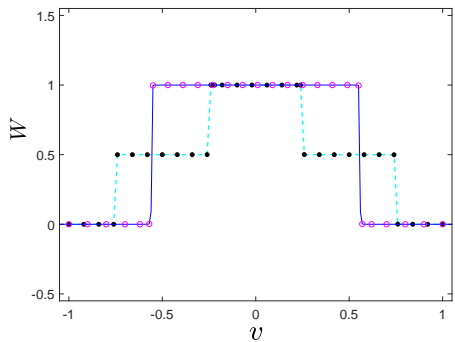}
\caption{(Color online)
Numerical results of both Bloch (denoted by black dots) and non-Bloch (denoted by magenta circles) topological invariant $W$ as a function of the intracell hopping $v$. The decay rate is $\gamma=1$ and the intercell hopping strength is $r=0.5$.
}
\label{fig7}
\end{figure}

Based on this Bloch Hamiltonian, winding numbers \cite{pupillo} under different values of $v$ are calculated which are denoted by black dots in Fig. \ref{fig7}. Unfortunately, the topologically nontrivial region revealed in Fig. \ref{fig7} doesn't match well the region in Figs. \ref{fig5} and \ref{fig6} where edge states emerge. And as shown in Fig. \ref{fig7}, the winding number has a fractional value of $1/2$ in two regions.

Therefore, we turn to resort to the so-called non-Bloch topological invariants\cite{shunyu,murakami}. A static rotation\cite{murakami} $e^{i \pi\sigma_x/4 } H'_k e^{-i \pi\sigma_x/4 }$ about $x$ axis maps $H'_k$ into $H''_k$, namely, $\sigma_x\!\!\rightarrow\!\!\sigma_x$, $\sigma_z\!\!\rightarrow\!\!\sigma_y$.
The non-Bloch Hamiltonian $H_{\beta}$ is obtained conveniently from $H''_k$ by implementing the replacement $e^{ik}\rightarrow \beta$, $e^{-ik}\rightarrow \beta^{-1}$. Explicitly, the non-Bloch Hamiltonian reads
\begin{eqnarray}
H_{\beta}=\left( v+\frac{\gamma}{4}+r\beta^{-1}\right) \sigma_{+} + \left( v-\frac{\gamma}{4}+r\beta\right) \sigma_{-},
\label{Hbeta}
\end{eqnarray}
in which $\sigma_{\pm}=\left( \sigma_x \pm i\sigma_y \right)/2$ and $\beta$ is defined on generalized Brillouin zone (GBZ)\cite{shunyu,murakami}. Correspondingly, the generalized Q matrix\cite{ryu} is defined as
\begin{eqnarray}
Q(\beta)=\left| \tilde{u}_R(\beta)\right> \left< \tilde{u}_L(\beta) \right| - \left| u_R(\beta)\right> \left< u_L(\beta) \right|
\label{Q}
\end{eqnarray}
where $\left| \tilde{u}_R(\beta)\right>\equiv\sigma_z \left|u_R(\beta)\right>$, $\left| \tilde{u}_L(\beta) \right> \equiv \sigma_z \left| u_L(\beta) \right>$, and
$\left|u_R(\beta)\right>$, $\left< u_L(\beta) \right|$ are the right and left eigenvector of $H_{\beta}$, respectively.
The Q matrix is off-diagonal and can be written in a form as $\left({\,}_{q^{-1}}{\;}^q\right)$. Accordingly, the non-Bloch winding number\cite{shunyu} is defined as
\begin{eqnarray}
W=\frac{i}{2\pi} \int_{GBZ} q^{-1}dq.
\end{eqnarray}
For the case with $r=0.5$ and decay strength $\gamma=1$, we numerically calculate the non-Bloch winding number $W$ as a function of the intracell hopping $v$. As shown in Fig. \ref{fig7}, it is clear that for $v \in \left[-0.559,0.559\right]$ the system is topological nontrivial with the non-Bloch winding number $W=1$. Comparing Fig. \ref{fig5}d and Fig. \ref{fig7} carefully, one can find that the edge modes in the single-particle energy spectrum could be well predicted by the non-Bloch topological invariant $W$.

\begin{figure}[b]
\includegraphics[width=0.46\textwidth]{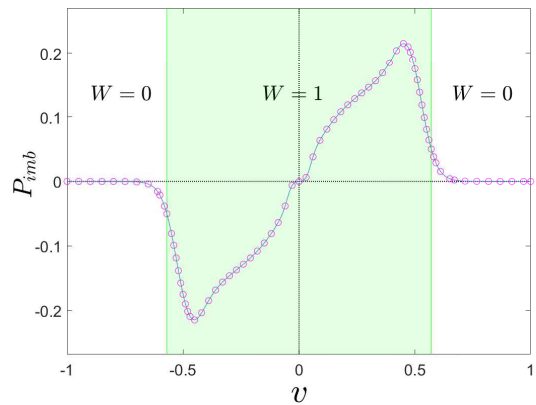}
\caption{(Color online)
Decay probability imbalance $P_{imb}$ between the two edge unit cells as a function of the intracell hopping $v$.
Region with the non-Bloch winding number $W=1$ is indicated by green-colored background.
The lattice size is $L=51$ (unit cell) with the decay rate $\gamma=1$ and intercell hopping $r=0.5$.
}
\label{fig8}
\end{figure}

Finally, we implement numerically the quantum walk on a finite bipartite non-Hermitian lattice with $L=51$ unit cells repeatedly with the intracell hopping $v$ scanning through the parametric region $\left[-1,1\right]$. The decay rate is set to be $\gamma=1$ and the intercell hopping is fixed at $r=0.5$. Based on various distributions of decay probability $P_m$ obtained during the numerical simulation above, we plot in Fig. \ref{fig8} the decay probability imbalance $P_{imb}$ between the two edge unit cells as a function of the intracell hopping $v$. Specifically, $P_{imb}$ is defined as
\begin{eqnarray}
P_{imb}=P_{l}-P_{r}
\end{eqnarray}
with $l$ and $r$ are the indices of the leftmost unit cell and the rightmost unit cell, respectively.
For convenience of comparison, different parametric regions with different non-Bloch winding numbers are indicated by different colors. Clearly as shown in Fig. \ref{fig8}, appearance of the counterintuitive distributions of local decay probability $P_{m}$ is intimately related to the topological nontrivial region with non-Bloch winding number $W=1$ except for tiny mismatches at edges of the region.
We infer that these tiny mismatches emerges as a result of finite-size effects since our study is concentrated on finite lattices. However, what we want to emphasize here is that the topological nontrivial region can be taken as a guide to tell us where it's possible to observe the intriguing distributions of local decay probability.
When the edge modes are located at the left edge unit cell (see Fig. \ref{fig5}c), conspicuous occupation of the local decay probability on the leftmost unit cell occurs. Similarly, when the edge modes are located on right edge unit cell (see Fig. \ref{fig5}a), impressive portion of the probability decays from the rightmost unit cell. Interestingly, it seems that the edge state has an attractive effect to the quantum walker walking on the non-Hermitian lattice. This is quite different from the case of Hermitian case\cite{liwang17}, in which edge state exhibits repulsive behavior to the quantum walker initiated in the bulk. When it comes to the case of zero intracell hopping, each of the two edge states is localized on one of the two edge unit cells, see Fig. \ref{fig5}b. The attractive effects of the two edge states seem to balance in power. Therefore, a almost symmetric distribution of the local decay probability comes into force, see Fig. \ref{fig4}.
Consistently, deep into parametric regions where the non-Bloch winding number $W$ valued zero, no edge states show up, see Figs. \ref{fig5} and \ref{fig6}. Therefore, as shown in Figs. \ref{fig2} and \ref{fig3}, the resultant distributions of local decay probability $P_m$ are asymmetric and back to normal.

\section{Conclusions} \label{summary}

In summary, we have investigated the single-particle continuous-time quantum walk on a finite bipartite non-Hermitian lattice with pure loss. Focusing on the resultant distribution of local decay probability $P_m$, intriguing phenomenon is found, in which impressive population of the decay probability appears on edge unit cell although it is the farthest from the starting point of the quantum walker. Detailed numerical simulations reveal that the intracell hopping $v$ of the lattice can be used to modulate the quantum walker to reach the leftmost unit cell, the rightmost unit cell or none of them with a relative high portion of the probability. We then investigate the energy spectrum of the non-Hermitian lattice under open boundary condition. Edge modes are shown existing in the real part of the energy spectrum. Basing on its mathematical connection to a similar model, we show that the edge modes are well predicted by a non-Bloch topological invariant. The occurrence of conspicuous population of the local decay probability on either edge unit cell is closely related to the existence of edge states and their specific properties.
The model could be experimentally realized with an array of coupled resonator optical waveguides along the line of Ref.\cite{taylor,lee}. The counterintuitive distributions shown in Figs. \ref{fig2} and \ref{fig3} should be observed experimentally. The dynamics of the quantum walker running on such a non-Hermitian lattice behaves quite like a quantum switch. The mechanism may have prosperous applications in the designing of microarchitectures for quantum information and quantum computing in future.

\section{Acknowledgements}
This work is supported by NSF of China under Grant Nos. 11404199 and 11674201, NSF for Shanxi Province Grant No. 1331KSC, NSF for youths of Shanxi Province No. 2015021012, and research initiation funds from SXU No. 216533801001.

\end{document}